\documentclass[aps,prd,twocolumn,superscriptaddress,nofootinbib,preprintnumbers]{revtex4-1}
\usepackage{amsmath}
\usepackage{graphicx}
\usepackage{subfigure}
\usepackage{amssymb}
\usepackage{xcolor}
\usepackage{multirow}
\usepackage{cancel}
\usepackage{color}
\usepackage{ulem}
\usepackage{listings}
\usepackage{xcolor}
\usepackage{float}
\usepackage{slashed}
\usepackage{wrapfig}
\usepackage{mathtools}
\usepackage[colorlinks,citecolor=blue]{hyperref}
\usepackage[T1]{fontenc}  
\usepackage[utf8]{inputenc}  
\usepackage{times}
\begin{document}
	
\title{Neutrinogenic CMB spectral distortions}

\author{Shao-Ping~Li}
\email{lisp@het.phys.sci.osaka-u.ac.jp}
\affiliation{Department of Physics, The University of Osaka, Toyonaka, Osaka 560-0043, Japan}
\author{Jens~Chluba}
\affiliation{Jodrell Bank Centre for Astrophysics, Alan Turing Building, University of Manchester, Manchester M13 9PL}
	
\begin{abstract}
Extra radiation injection  after neutrino decoupling in the early Universe contributes to  the  effective  number of neutrino species   that can be constrained by the cosmic microwave background (CMB).  However, any effective neutrino number itself cannot uniquely determine the underlying source. We argue that the degeneracy can be relaxed by CMB spectral distortions, which are caused by energy exchange between  the  extra radiation and photons. We consider neutrinogenic CMB spectral distortions, where extra energy is released in the form of neutrinos but still creates  the CMB spectral  distortions via electroweak interactions.  The synergy between the effective neutrino number and CMB spectral distortions provides a complementary  probe of hidden sectors  that dominantly couple to neutrinos, opening up parameter space that can be targeted by joint CMB anisotropy and spectral distortion experiments.
\end{abstract}
	
\maketitle

\preprint{OU-HET 1292} 
	
\section{Introduction}	\label{sec:intro}
 Big-bang nucleosynthesis (BBN) and the cosmic microwave background (CMB) are two well established probes in cosmology, where observations in the past few years have confirmed the picture of the early Universe to an unprecedented precision \citep{Planck:2018vyg}. Both BBN and CMB are sensitive to the Hubble expansion, corresponding to cosmic temperatures at MeV and eV scales, respectively. This sensitivity is mediated by the effective number of neutrino species, $N_{\rm eff}$, which yields $N_{\rm eff}=3.043-3.046$~\cite{Mangano:2001iu,Mangano:2005cc,deSalas:2016ztq,EscuderoAbenza:2020cmq,Akita:2020szl,Froustey:2020mcq,Bennett:2020zkv,Cielo:2023bqp} in the Standard Model (SM) of particle physics  after neutrinos fully decouple from the SM plasma at around 10~keV temperatures. Current  observational accuracy can constrain shifts of $\Delta N_{\rm eff}$ at $\mathcal{O}(0.1)$~\cite{Planck:2018vyg,Fields:2019pfx,DESI:2024mwx,SPT-3G:2024atg,ACT:2025tim,DESI:2025ejh}, while upcoming CMB and BBN experiments promise to reach  $\mathcal{O}(0.01)$ accuracy~\cite{EUCLID:2011zbd,CMB-S4:2016ple,SimonsObservatory:2018koc,Abazajian:2019eic,Sehgal:2019ewc,Yeh:2022heq,Euclid:2024imf}. 

Being a key observable in the early Universe, $N_{\rm eff}$  has been widely applied to probe various physics processes that modify the Hubble expansion during neutrino decoupling, BBN and the subsequent CMB epochs. This encompasses a broad class of cosmological scenarios beyond the SM, motivated by the unsolved problems related to neutrino masses, dark matter, and dark energy, among other things. Nevertheless, it has been known that any deviation of $N_{\rm eff}$, if detected, cannot uniquely determine the underlying source, since numerous extra energy injection as SM left-handed neutrinos, right-handed Dirac neutrinos, some hidden radiation such as dark photons, majorons, and axionlike particles, as well as gravitational waves, can readily give rise to the same prediction of $N_{\rm eff}$. Therefore, promoting $N_{\rm eff}$ as a powerful observable to discriminate different particle physics origins necessitates breaking inevitable degeneracies~\cite{Baumann:2015rya}.

CMB spectral distortions describe the departure of the background photon spectrum from a perfect blackbody distribution. These signals can be created in the form of the $\mu$ distortion at a high redshift $z\simeq 2\times 10^6$ or as the $y$-distortion below $z=5\times 10^4$~\citep{Burigana:1991eub,Hu:1992dc}. The COBE/FIRAS experiments showed that the $\mu$ and $y$ distortions are restricted to $|\mu|<9\times 10^{-5}, |y|<1.5\times 10^{-5}$~\cite{Mather:1993ij,Fixsen:1996nj}, and recently the reanalysis of the FIRAS data improved the bounds by about a factor of two, yielding $|\mu|<4.7\times 10^{-5}$~\cite{Bianchini:2022dqh}, $|y|<8.3\times 10^{-6}$~\cite{Sabyr:2025hwd}. Several standard sources in the $\Lambda$CDM model such as the Silk damping~\cite{Hu:1992dc,Chluba:2012gq} and the thermal Sunyaev-Zeldovich effect~\cite{ZS1969,ZS1970,Refregier:2000xz,DeZotti:2015awh}, as well as well-motivated cosmic scenarios such as relic hidden particle decay~\cite{Hu:1993gc,Chluba:2016bvg}, dark matter annihilation~\cite{McDonald:2000bk,Chluba:2011hw,Chluba:2013wsa,Li:2024xlr} and black hole evaporation~\cite{Carr:2009jm,Nakama:2017xvq} can induce the $\mu$ and $y$ distortions beyond the current COBE/FIRAS  detection limits. In the past decades, however, most studies of CMB spectral distortions from exotic sources have focused on direct photon energy release, while energy release to non-electromagnetic species like neutrinos or neutral dark radiation is commonly treated via $N_{\rm eff}$ without connections to CMB spectral distortions.

While there is a three-decade time gap after the COBE/FIRAS measurements, it is exciting news that several mission programs are now in the implementation and active concept development phase, hopefully being set to commence first observations in a few years from now. These include ground-based experiments such as COSMO~\cite{Masi:2021azs} and TMS~\cite{TMS}, balloon-borne experiments such as BISOU~\cite{Maffei:2021xur}, and space-based concepts such as {\it PIXIE}~\cite{Kogut:2011xw,Kogut:2019vqh,Kogut:2024vbi}, {\it FOSSIL}~\cite{fossil2022}, and {\it SPECTER}~\cite{Sabyr:2024lgg}, the latter of which could reach the spectral distortions down to $|\mu|\sim 10^{-8}, |y|\sim10^{-9}$.

In this work, we consider breaking the degeneracy of $N_{\rm eff}$ via the correlation between $N_{\rm eff}$ and CMB spectral distortions. The simple physics that forms the correlation is the conversion between the injected radiation and the background photons. This may have already been inherited from the same underlying physics that yields the radiation injection, and hence could be a general phenomenon across a wide class of particle physics scenarios. Given that the detection sensitivities of CMB spectral distortions are much higher than that of $N_{\rm eff}$, it is possible that the extra energy release causing a shift of $N_{\rm eff}$ has a small leakage into photon energy and thereby creates observable CMB spectral distortions. 

Recently, CMB spectral distortions due to electroweak cascades of ultrahigh-energy neutrino injection received some attention~\cite{Acharya:2020gfh,Hambye:2021moy,Bianco:2025boy}. For neutrino injection with energies above 100~GeV, current constraints from CMB spectral distortions and BBN photo- or hadro-disintegration become more severe, which largely excludes an observable shift of  $\Delta N_{\rm eff}$ as well as joint observations of $\Delta N_{\rm eff}$ and CMB spectral distortions. On the other hand, joint observations of $\Delta N_{\rm eff}$ and the CMB spectral distortions may also be realized by photon energy release~\cite{Chluba:2020oip}. This occurs via large energy release to photons not long before $z\simeq 2\times 10^6$, where a negative shift of $\Delta N_{\rm eff}$ is created due to the shift of the photon temperature, and a small CMB $\mu$ distortion can be induced afterwards by exponentially diluting the large energy release \cite{Chluba:2020oip, Acharya:2021zhq}.  
 
Instead of a direct modification to the background photons, we demonstrate that if the extra energy release is in the form of neutrinos, a large parameter space can still create a correlation between $N_{\rm eff}$ and CMB spectral distortions across a wide cosmic epoch, even if the injected neutrino energy is as low as a few GeV. This generic synergy can also be applied to any non-electromagnetic radiation that has indirect couplings to electromagnetic species. 

\section{Indirect photon heating from neutrino annihilation}
Even if extra energy injection in the early Universe is exclusively through the neutrino channel, 
the injected neutrino energy can still partly transfer to the electromagnetic plasma within the SM interactions. 
For example, if  $m_e^2/T\lesssim E_\nu \lesssim m_\mu^2/T$, where $E_\nu$ is the monochromatic neutrino energy injected while  $m_e, m_\mu$ denote the electron and muon masses, respectively,  the conversion can proceed via tree-level coannihilation to electron-positron pairs, $\nu_{\rm inj}+\bar\nu_{\rm bg} (\bar\nu_{\rm inj}+\nu_{\rm bg})\to e^++e^-$. Here, $\nu_{\rm inj}$ denotes injected neutrinos, while thermal background neutrinos are indicated by $\nu_{\rm bg}$. It implies that a CMB spectral distortion that can be formed at around 0.5~keV requires $E_\nu\gtrsim 500$~MeV. While the electron-positron pairs can still be produced in coannihilation if $E_\nu<500$~MeV, the production rate will be suppressed by the Boltzmann exponential factor from the high-energy thermal tail of the background neutrino distribution.  
 
For $E_\nu\gtrsim m_\mu^2/T$, additional channels will open to produce heavier charged leptons, mesons and even gauge bosons. In the heavy-mass regime $E_\nu>100$~GeV, the precise energy distribution necessitates a detailed simulation of the cascade processes, as studied in Refs.~\cite{Cirelli:2010xx,Acharya:2020gfh,Bianco:2025boy}. In particular, any monochromatic neutrino spectrum will be modified due to secondary production of low-energy neutrinos from electroweak gauge boson cascades. The full numerical analysis in this heavy-mass regime is beyond the scope of this work. 
 
On the other hand, for lower $E_\nu$, in particular, $1~\text{MeV}\lesssim E_\nu\lesssim 100$~MeV, coannihilation is  kinematically forbidden. However, the pair annihilation channel  $\nu_{\rm inj}+\bar \nu_{\rm inj} \to e^++e^-$ remains open to create the electron-positron pairs. One can expect that there would be a threshold $E_\nu$ at which pair annihilation starts to dominate the energy transfer, warranting comprehensive investigation for the accurate transfer rate between neutrinos and electromagnetism.
 
The produced electron-positron pairs will undergo rapid repeated inverse Compton scattering  $e^\pm +\gamma_{\rm bg}\to e^\pm+\gamma$, where $\gamma_{\rm bg}$ is  the background photon. Consequently, the background  photons will be heated due to the energy loss of electrons and positrons. If  the electron-positron pairs are relativistic, the dominant energy loss  of $e^\pm$ pairs  is via inverse Compton scattering~\cite{Blumenthal:1970gc,Chen:2003gz,Padmanabhan:2005es}, which is much faster than Hubble expansion   before recombination.  The heated photons will subsequently interact with the background plasma,  redistributing the extra energy via Compton scattering $\gamma+e_{\rm bg}\to \gamma +e$ and participating  in fast photon-number changing processes (double Compton scattering and bremsstrahlung) in the low-energy band. This gives rise to departures from the perfect blackbody distribution in the form of the $\mu$, $y$, or residual distortions, depending on the conversion time~\cite{Illarionov1975,Chluba:2011hw, Khatri:2012tw,Chluba:2013vsa,Acharya:2018iwh}.

Given the above discussions, we are to calculate the energy leakage from the injected neutrinos to electron-positron pairs via coannihilation and pair annihilation. Note that  injected neutrinos scattering off background electrons is suppressed since the electron number density is smaller than the background photons by a factor of $10^{-10}$, i.e., the small baryon-to-photon ratio. Similarly, the pair annihilation rate between injected neutrinos is suppressed, with some gains from the energy-dependence of the cross section, as we will show here.

\section{$\boldsymbol {N_{\rm eff}}$ meets the $\boldsymbol \mu$ distortion}
Throughout we will focus on the connection between $N_{\rm eff}$ and the CMB $\mu$ distortion, which, to a good approximation, allows us to perform semi-analytic analysis. Applications to the $y$ distortion and residual distortions during the transition epoch from the $\mu$ to the $y$ distortions are straightforward, but necessitate more dedicated numerical calculations. 

For simplicity, we assume that the injected neutrinos are in a single flavor. Due to fast  neutrino oscillations after neutrino decoupling, the single neutrino flavor will convert to other flavors with approximately an equal distribution for each flavor. The conversion channels come from $\nu_i+\bar \nu_j\to e^++e^-$ induced  by neutral and charged weak currents, where $i,j=1,2,3$ denote the mass eigenstates of neutrinos.  Owing to the structure of the neutrino PMNS mixing matrix~\cite{Esteban:2024eli}, we simply do the summation of $i,j$ over three mass eigenstates by multiplying the electron-neutrino annihilation rate $\nu_e+\bar\nu_e\to e^++e^-$ by a factor of 3. On the other hand, we will also neglect the final-state radiation that directly produces photons, $\nu+\bar \nu\to e^-+e^-+\gamma$, which is of higher order in the electromagnetic coupling. Such processes can receive a certain logarithmic enhancement in the infrared regime, but the contribution to CMB spectral distortions from soft photons will be small before recombination~\cite{McDonald:2000bk}. While our approximate treatments may lead to theoretical uncertainties of the annihilation rate at $\mathcal{O}(10\%)$, the main conclusions drawn in the following would not be changed. 
 
 The energy transfer rate into the electron-positron pair is then given by 
\begin{align}\label{eq:drhoe/dt}
	\frac{{\rm d}\rho_{e\bar e}}{{\rm d}t}=\int {\rm d}\Pi  \,\tilde \delta^4(p)|\mathcal{M}|^2_{\nu\to e} (E_e+E_{\bar e})f_\nu f_{\bar \nu_{}}\,,
\end{align}
where $\tilde \delta^4(p)\equiv(2\pi)^4 \delta^4\left(p_\nu+p_{\bar \nu_{}}- p_e-p_{\bar e}\right)$ with $p_i$ the four-momentum, and  the phase-space factor is defined as
\begin{align}
{\rm d}\Pi\equiv \prod_{i=\nu,\bar \nu, e,\bar e}\frac{{\rm d}^3 {\bf p}_i}{(2\pi)^3 2E_i}\,, 
\end{align}
with ${\bf p_i}$ the spatial momentum, and we have neglected the Pauli-blocking effect from the final-state electron-positron pairs. With  $E_e+E_{\bar e}=E_\nu+E_{\bar \nu}$, we can calculate the electron-positron phase-space integration in the center-of-mass frame as
\begin{align}\label{eq:ee-int}
	\int {\rm d}\Pi_e {\rm d}\Pi_{\bar e} \,\tilde \delta^4 \left(p\right)|\mathcal{M}|^2_{\nu\to e}
=\frac{G_{\rm F}^2 s^2}{3\pi}\mathcal{G}\theta(s-4m_e^2)\,,
\end{align}
where $G_{\rm F}=1.167\times 10^{-5}~\text{GeV}^{-2}$ is the Fermi constant,  $s=(p_\nu+p_{\bar \nu})^2$, and $\theta(s-4m_e^2)$ is the Heaviside step-function. The function $\mathcal{G}$ is defined by 
\begin{align}\label{eq:G}
	\mathcal{G}\equiv \left(1+\frac{2m_e^2}{s}\right)g_W-\frac{3m_e^2}{s}\,,
\end{align}
where $g_W\equiv \left(1+4\sin\theta_W^2+8\sin\theta_W^4\right)$, and $\theta_W$ is  the weak mixing angle with  $\sin\theta_W^2\approx 0.23$.  This result is consistent with the expression for electron-neutrino annihilation $\nu_e+\bar\nu_e\to e^++e^-$~\cite{Hannestad:1995rs}. 

The electron-positron kinetic energy that will heat the background photons is  approximately  given by 
\begin{align}\label{eq:drho_gamma/dt}
	\frac{{\rm d}\rho_\gamma}{{\rm d}t}\approx \frac{{\rm d}\rho_{e\bar e}}{{\rm d}t}\,.
\end{align}
While not all the energy stored in the electron-positron pairs will convert into photon energy,  Eq.~\eqref{eq:drho_gamma/dt} is a good approximation to predict the $\mu$-distortion formation for relativistic $e^\pm$~\cite{Blumenthal:1970gc,Chen:2003gz,Padmanabhan:2005es,Li:2024xlr}. The generated $\mu$ distortion can be calculated  by integrating the energy transfer rate over the entire production history~\cite{Sunyaev:1970er, Chluba:2013vsa,Chluba:2016bvg,Li:2024xlr}
\begin{align}\label{eq:mu-dis}
	\mu\approx 1.4 \int_{0}^{\infty} \mathcal{J}_\mu(T)\frac{{\rm d}\rho_\gamma/{\rm d}t}{\rho_\gamma HT}{\rm d}T\,.
\end{align}
Here, we have used the cosmic temperature as the time variable via ${\rm d}T/{\rm d}t=-H T$, where $H\approx 1.66\sqrt{g_\rho}T^2/M_{\rm P}$ is the Hubble parameter with  $M_{\rm P}\approx 1.22\times 10^{19}$~GeV the Planck mass and $g_\rho(T)$ the effective degrees of freedom in energy density. $\mathcal{J}_\mu(T)$ is the visibility function for the epoch during which the $\mu$ distortion is formed~\cite{Burigana:1991eub,Hu:1992dc,Chluba:2013kua,Chluba:2016bvg}
\begin{align}\label{eq:mu-vis-2}
\mathcal{J}_\mu(T)=e^{-(T/T_\mu)^{5/2}}\theta(T-T_{\mu y})\,,
\end{align}
with $T_\mu\approx 0.47$~keV (corresponding to $z=2\times 10^6$) being the highest temperature for the $\mu$ distortion formation and $T_{\mu y}\approx 12$~eV (corresponding to $z=5\times 10^4$) being the transition temperature from the $\mu$ to the $y$ distortion formation. 

Before calculating the $\mu$ distortion, let us make a qualitative and fast estimate about the electromagnetic energy leakage once  $\Delta N_{\rm eff}$ is created by the neutrino injection, where the current COBE/FIRAS measurements correspond to a bound of $\Delta \rho_\gamma/\rho_\gamma\lesssim 6\times 10^{-5}$ at $95\%$ confidence level while the future target on the $\mu$ and $y$ distortions can reach $\Delta \rho_\gamma/\rho_\gamma\sim 10^{-8}-10^{-9}$. Here $\Delta \rho_\gamma$ denotes the total electromagnetic energy leakage and $\rho_\gamma\approx 0.66T^4$ is the background photon energy density. This simple approximation of estimating the electromagnetic leakage will also allow one to further anticipate the correlation between $\Delta N_{\rm eff}$ and the $y$ distortion.

The $N_{\rm eff}$ shift is defined as
\begin{align}\label{eq:DeltaNeff-nu}
	\Delta N_{\rm eff}=\frac{8}{7}\left(\frac{11}{4}\right)^{4/3} \left(\frac{ \Delta \rho_\nu}{\rho_\gamma}\right)\,,
\end{align}
where $\Delta \rho_\nu$ is the non-thermal neutrino energy release after neutrino decoupling. We can estimate the normalized energy transfer by integrating Eq.~\eqref{eq:drho_gamma/dt} over the temperature (redshift) history during which the CMB spectral distortions are formed. To obtain the relation between the electromagnetic energy transfer and $\Delta N_{\rm eff}$, let us assume that the neutrino energy release is already complete before the dawn of the CMB spectral distortion formation, such that $\Delta N_{\rm eff}$  is a constant after $T= 0.47$~keV. 

When the energy transfer is induced by neutrino coannihilation $\nu_{\rm inj}+\bar \nu_{\rm bg}\to e^++e^-$, we have
\begin{align}
    \frac{\Delta \rho_{\gamma}}{\rho_\gamma}
    \simeq 10^{-9}\left(\frac{E_\nu}{1~\text{GeV}}\right)\left(\frac{\Delta N_{\rm eff}}{0.05}\right)\left(\frac{T_i}{0.1~\text{keV}}\right)^2.\label{eq:Deltarho_gamma-rho_gamma-1}
\end{align}
The detailed derivation that leads to the above semi-analytic result is presented in Appendix~\ref{app:rate}.  Here, $T_i$ and $T_f$ denote the initial and final moments of the injection, respectively, and we take the limit $T_i\gg T_f$ for simplicity. In general,  $E_\nu\gtrsim 500$~MeV is expected to generate electron-positron pairs from coannihilation, but we have neglected the redshift effects on the injected neutrino energy $E_\nu$ and on the energy spectrum for illustration purpose. These approximations only hold if most of the injection events are complete not long before $T_i$ such that redshift effects can be neglected. Then, we can infer that at the $\mu$ or $y$ era Eq.~\eqref{eq:Deltarho_gamma-rho_gamma-1} yields
\begin{align}\label{eq:Drho-co-muy}
     \frac{\Delta \rho_{\gamma,\rm co}}{\rho_\gamma}&\simeq c_i \left(\frac{E_\nu}{1~\text{GeV}}\right)\left(\frac{\Delta N_{\rm eff}}{0.1}\right),
\end{align}
where $c_i= 5\times 10^{-8}$ for the $\mu$ distortion with $T_i=T_\mu$  and $c_i= 3\times 10^{-11}$  for the $y$ distortion with $T_i=T_{\mu y}$.  It implies that for $\Delta N_{\rm eff}\sim 0.1$, the current bound of the $\mu$ distortion will supersede that of $N_{\rm eff}$ for  ultrahigh-energy neutrino injection $E_\nu\gtrsim 10^3$~GeV. On the other hand, it indicates that joint observations of $\Delta N_{\rm eff}\sim 0.1$ and $\mu\sim 10^{-8}$ can be reached for $1~\text{GeV}\lesssim E_\nu\lesssim 10^3$~GeV, and  joint observations of $\Delta N_{\rm eff}\sim 0.1$ and $y\sim 10^{-9}$ can be reached for $ E_\nu\gtrsim 10^2$~GeV. Note that, however, when $E_\nu$ is above the electroweak scale, the electroweak gauge boson cascades become important to modify the neutrino energy spectrum and the total electromagnetic energy leakage, and hence the resulting $\mu$ and $y$ distortions could differ significantly. 

When the electromagnetic energy transfer is induced by neutrino pair annihilation, $\nu_{\rm inj}+\bar \nu_{\rm inj}\to e^++e^-$, we have
\begin{align}
    \frac{\Delta \rho_{\gamma}}{\rho_\gamma}\simeq 10^{-10}& \left(\frac{E_\nu}{1~\text{GeV}}\right)
 \left(\frac{\Delta N_{\rm eff}}{0.1}\right)^2\left(\frac{T_i}{0.1~\text{keV}}\right)^2,\label{eq:Deltarho_gamma-rho_gamma-2}
\end{align}
where similar approximations to derive Eq.~\eqref{eq:Deltarho_gamma-rho_gamma-1} were also applied here; See Appendix~\ref{app:rate} for more details. It yields the energy transfer
\begin{align}\label{eq:Drho-pair-muy}
     \frac{\Delta \rho_{\gamma,\rm pair}}{\rho_\gamma}\simeq c_i \left(\frac{E_\nu}{1~\text{GeV}}\right)\left(\frac{\Delta N_{\rm eff}}{0.1}\right)^2,
\end{align}
with $c_i=1.4\times 10^{-9}$ for the $\mu$ distortion at $T_i=T_\mu$ and $c_i=8.8\times 10^{-13}$  for the $y$ distortion at $T_i=T_{\mu y}$. It then  implies that  if   $\Delta N_{\rm eff}\sim 0.1$ is created,  the $\mu (y)$ distortion  residing  within the detection limit $\mu\sim 10^{-8}(y\sim 10^{-9})$ can  be generated simultaneously via neutrino pair annihilation with $E_\nu\gtrsim 1 (10^3)$~GeV. 

The coefficient from Eq.~\eqref{eq:Drho-pair-muy} is one order-of-magnitude smaller than that from Eq.~\eqref{eq:Drho-co-muy}, suggesting that the neutrino energy from pair annihilation should not be too small for the electromagnetic energy leakage becoming comparable with coannihilation. Indeed,
we see that the minimal $E_\nu$ that can create observable CMB spectral distortions is similar in both cases. This suggests a fact that significant effects on the CMB spectral distortions from pair annihilation should be essentially achieved by increasing the neutrino energy to those values at which coannihilation already came into play. 

One might naively expect that increasing $E_\nu$ in Eq.~\eqref{eq:Deltarho_gamma-rho_gamma-1} and Eq.~\eqref{eq:Deltarho_gamma-rho_gamma-2} can enhance the CMB spectral distortions. However, $\Delta N_{\rm eff}$ will also be enhanced and could exceed the observational bounds, as it depends linearly on the injected neutrino energy.  More detailed calculations presented below confirm these expectations.  In particular, MeV-scale neutrino injection will lead to a large $N_{\rm eff}$ beyond the current bounds before the associated CMB $\mu$ distortion can reach future detection limits, rendering pair annihilation ineffective as a whole for the electromagnetic energy leakage below $E_\nu=1$~GeV.

To obtain a more precise prediction of the $\mu$ distortion, we need to take into account the redshift effects, and complete the calculation of Eq.~\eqref{eq:drhoe/dt} by specifying the neutrino distribution function, both of which depend on the injection source and injection time. In the following, we present the typical example from long-lived particle decay, but generalization to any neutrino injection is straightforward.

\section{An example: long-lived particle decay}	\label{sec:neutrino-release}
Here, we apply the generic synergy of $N_{\rm eff}$ and the $\mu$ distortion to long-lived particles that were once present in the early Universe but with a lifetime shorter than the age of the current Universe. 
For a model-independent analysis, we parameterize the abundance of the long-lived particle $X$ as
\begin{align}\label{eq:YX-def}
		Y_X\equiv \frac{n_X}{s}\,,
\end{align}
where $s=2\pi^2 g_s(T)T^3/45$ denotes the entropy density with the effective degrees freedom  $g_s(T)\approx 3.34$ after neutrino decoupling. The above parameterization works as the initial condition for $Y_X$, which is applicable to the regime after $X$ decouples from the thermal plasma but prior to $X$ decay. While we turn agnostic on the particle physics origins, the following analysis can be applied directly in a given particle physics framework, or be generalized to include additional channels of electromagnetic energy leakage. For example, a long-lived neutral gauge boson $Z'$ itself may provide the source for neutrino injection~\cite{Li:2023puz}, or may induce additional neutrino annihilation $\nu+\bar\nu\to Z'\to e^++e^-$, where a small mass of $Z'$ can compensate for the suppression from small couplings to SM fermions and hence may lead to a larger annihilation rate than the SM prediction.

For neutrino energy release from long-lived particle decay, we can solve the Boltzmann equation for the neutrino distribution from nonrelativistic $X$ decay: $X\to \nu+\bar\nu$, which reads
\begin{align}\label{eq:Boltzmann-nu}
	\frac{\partial f_\nu}{\partial t}-H |{\bf p}_\nu| \frac{\partial f_\nu}{\partial |{\bf p}_\nu|}=\frac{8\pi \alpha m_X \Gamma_X}{2E_\nu}\int {\rm d}\Pi \tilde \delta^4(p) f_X\,,
\end{align}
where $\tilde \delta^4(p)\equiv(2\pi)^4 \delta^4\left(p_X-p_\nu-p_{\bar \nu_{}}\right)$, $\alpha$ denotes the branching ratio to neutrinos with $\Gamma_X=1/\tau_X$ being the total decay width, and 
\begin{align}
{\rm d}\Pi\equiv \prod_{i=X,\bar \nu}\frac{{\rm d}^3 {\bf p}_i}{(2\pi)^3 2E_i}\,.
\end{align}
By defining the dimensionless variables, 
\begin{align}
x\equiv \frac{T_{\nu\rm dec}}{T}\,,\quad  r=\frac{|{\bf p}_\nu|}{T}\,, \quad r_X\equiv \frac{m_X}{T_{\nu\rm dec}}\,,
\end{align}
with $T_{\nu\rm dec}$ the neutrino decoupling temperature, the solution to Eq.~\eqref{eq:Boltzmann-nu} with the initial condition $x_0=1$ can be analytically derived as
\begin{align}\label{eq:fnu_sol}
	f_\nu=\frac{16\pi^2 c_s \alpha \eta Y_X}{r r_X^2}e^{\eta (1-4r^2/r_X^2)}\Theta(r,r_X,x)\,,
\end{align}
where  $c_s$ is defined through the entropy density $c_s\equiv 2\pi^2 g_s/45\approx 1.47$,  $M^*_{\rm P}=0.6 M_{\rm P}/\sqrt{g_\rho}$, and  $g_s\approx g_\rho\approx 3.34$.  $Y_{X}$ is the initial $X$ number density yield  defined at the end of the neutrino decoupling temperature $T_{\nu\rm dec}$, as given by Eq.~\eqref{eq:YX-def}. For definiteness, we set $\alpha=1$ and $T_{\nu\rm dec}=50$~keV throughout the numerical analysis. The Heaviside step-function $\Theta(r,r_X,x)\equiv \theta(2r-r_X)\theta(x r_X-2r)$ reflects the monochromatic energy injection after $X$ becomes nonrelativistic, as well as the redshift effect of the neutrino energy after decay. The dimensionless parameter $\eta$ is defined through
\begin{align}
\eta\equiv \frac{\Gamma_X M^*_{\rm P}}{2T_{\nu\rm dec}^2}\,,
\end{align}
characterizing the ratio of the decay rate to the Hubble expansion rate at $T_{\nu\rm dec}$.

\begin{figure}[t]
	\centering
	\includegraphics[scale=0.5]{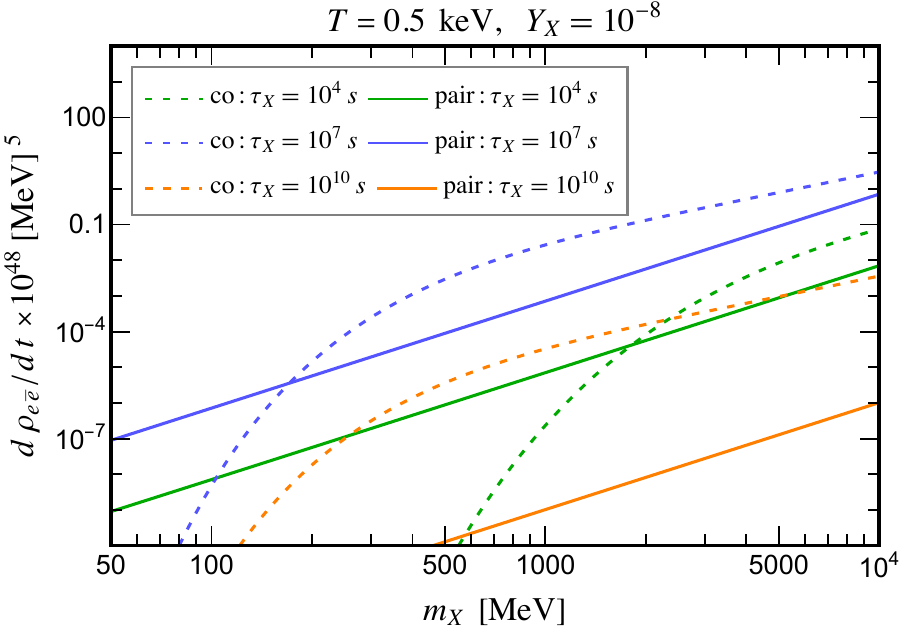} 
	\caption{\label{fig:ratecomp} The electromagnetic energy transfer rates from coannihilation (co) and pair annihilation (pair) at $T=0.5$~keV with an initial yield $Y_X=10^{-8}$. Three lifetime examples $\tau_X=10^4, 10^7, 10^{10}$~s are chosen such that the long-lived particle decays  before, around, and near the end of  the $\mu$ distortion formation. 
	}
\end{figure}

The Boltzmann equation assumes that scattering effects on late-time evolution of the injected neutrinos are suppressed. This is indeed the case if the injected neutrino energy is not too high. We can justify this by considering the typical timescale predicted by coannihilation. Using Eq.~\eqref{eq:drhoe/dt} normalized to the injected neutrino energy density and then comparing to the Hubble expansion rate, one can check that for $T<T_{\nu\rm dec}$, the scattering timescale for the injected neutrinos with the background plasma would be shorter than the Hubble time if $E_\nu\gtrsim 10^3$~GeV. Therefore, for neutrino energy injection after neutrino decoupling and  below the electroweak scale,  the Hubble expansion is always faster than scattering, which implies that Eq.~\eqref{eq:Boltzmann-nu} is a good approximation to determine the injected neutrino energy spectrum.

\begin{figure*}[t]
	\centering
	\includegraphics[scale=0.43]{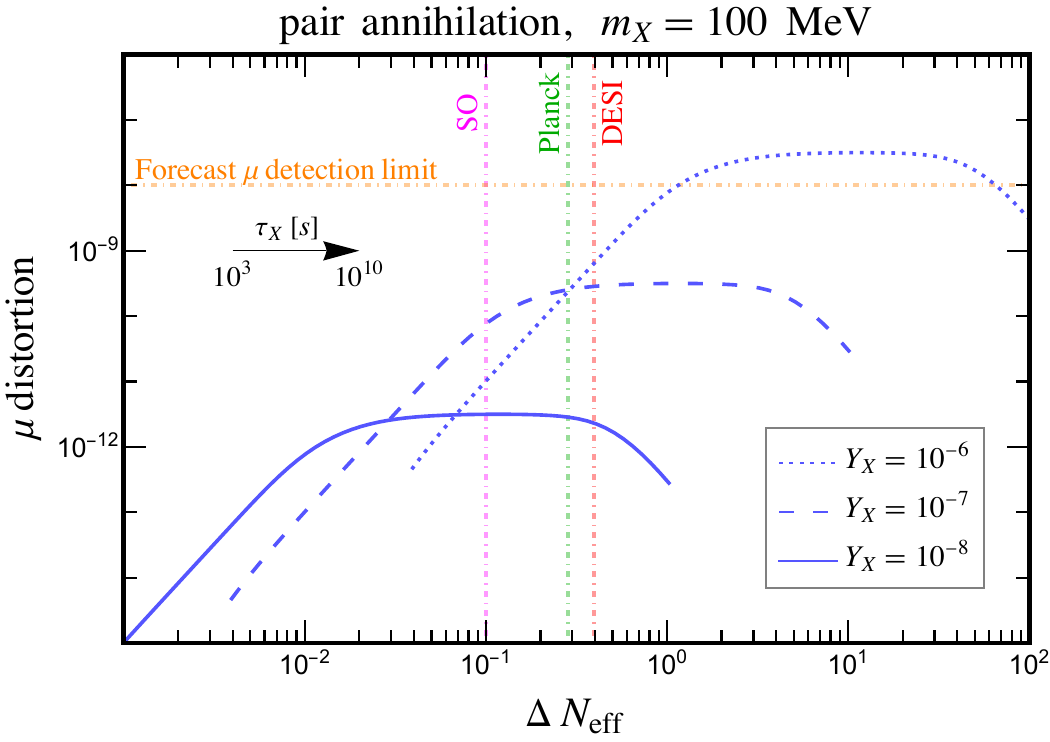} \quad 	\includegraphics[scale=0.43]{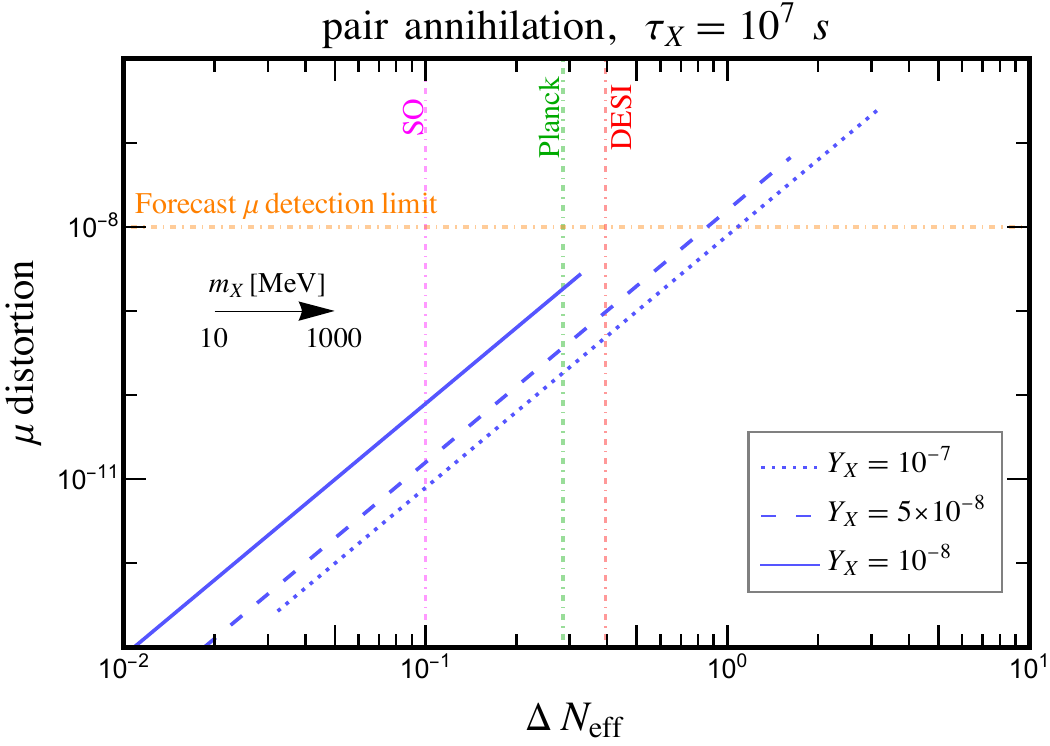}
	\caption{\label{fig:Neffmu-pair} The correlated predictions of $\Delta N_{\rm eff}$ and the CMB $\mu$ distortion from pair annihilation of injected neutrinos $\nu_{\rm inj}+\bar\nu_{\rm inj}\to e^++e^-$. The arrow in the left (right) panel denotes  the increase of  lifetime (mass). The current $2\sigma$ observational bounds from DESI~\cite{DESI:2024mwx}: $\Delta N_{\rm eff}<0.395$, and Planck~\cite{Planck:2018vyg}: $\Delta N_{\rm eff}<0.285$ are shown, together with the detection sensitivity at  $2\sigma$ significance level $\Delta N_{\rm eff}=0.1$ from upcoming Simons Observatory (SO) experiment~\cite{SimonsObservatory:2018koc}. We take $\mu=10^{-8}$ as the forecast detection limit of FOSSIL for reference, which is a factor of 2 smaller than the largest $\mu$ distortion predicted in the standard $\Lambda$CDM model~\cite{Chluba:2012gq,Chluba:2013wsa}.
	}
\end{figure*}

To determine the electromagnetic energy transfer from coannihilation, we substitute Eq.~\eqref{eq:fnu_sol} into Eq.~\eqref{eq:drhoe/dt} for $f_\nu$ and take the thermal distribution for the background antineutrino 
\begin{align}
	f_{\bar\nu}(E_{\bar \nu})=\frac{1}{e^{E_{\bar\nu}/T_\nu}+1}\,,
\end{align}
with $T_\nu=(4/11)^{1/3}T$ the neutrino temperature. In addition, we multiply Eq.~\eqref{eq:drhoe/dt} by a factor of 2 to include the conjugated process $\bar\nu_{\rm inj}+\nu_{\rm bg}\to e^++e^-$. For pair annihilation, we substitute Eq.~\eqref{eq:fnu_sol} into Eq.~\eqref{eq:drhoe/dt} for both $f_\nu$  and $f_{\bar \nu}$. 

We numerically  calculate  the energy transfer rates from coannihilation and pair annihilation, and show the results in Fig.~\ref{fig:ratecomp} by taking the   typical lifetimes  before, during, and after $z\simeq 2\times 10^6$. For coannihilation with $\tau_X\gtrsim 10^7$~s, we see that the  energy transfer rates become suppressed after $m_X\lesssim 500$~MeV, which is caused by the energy threshold of $e^\pm$ production. If  $\tau_X< 10^7$~s, the electromagnetic energy transfer would be  induced by injected  neutrinos with energy redshifting  from the earlier  epoch. Therefore, we see that for $\tau_X=10^4$~s,  $m_X$ (and hence $E_\nu$) should be larger to reach the energy threshold of $e^\pm$ production, and hence the curve starts to decrease already at a larger $m_X$.

In general, the pair annihilation rate becomes larger than the coannihilation rate when $m_X$  is much  below the coannihilation energy threshold of the  $e^\pm$ production. Above the energy threshold, the coannihilation rate dominates over the pair annihilation rate for  three typical lifetimes shown in Fig.~\ref{fig:ratecomp}. Nevertheless, given that ${\rm d} \rho_{e\bar e, \rm co}/{\rm d}t\propto Y_X$ while ${\rm d} \rho_{e\bar e, \rm pair}/{\rm d}t\propto Y_X^2$,  the pair annihilation channel can continue to be more important than  coannihilation above the energy threshold if  $Y_X$ is larger. This may also be inferred from Ref.~\cite{Bianco:2025boy} that considered $E_\nu\gg 1$~GeV, where $Y_X\gg 10^{-8}$ would have a significant impact via pair annihilation on disintegrating light nuclei formed during BBN.

At small $m_X$, Fig.~\ref{fig:ratecomp} indicates that  pair annihilation  will be the dominant channel for the electromagnetic energy leakage. Nevertheless, it remains to be seen if a significant $\mu$ distortion can be formed without creating a too large $\Delta N_{\rm eff}$. To see this, we calculate $\Delta N_{\rm eff}$ via Eq.~\eqref{eq:DeltaNeff-nu} by taking the asymptotic temperature at recombination $T\simeq 0.1$~eV, noting that $\Delta N_{\rm eff}$ will quickly become a constant after decay. Meanwhile, we  calculate the $\mu$ distortion from Eq.~\eqref{eq:mu-dis} with the same parameter set. 

We show the result from pair annihilation  in Fig.~\ref{fig:Neffmu-pair}, where the correlation of $\Delta N_{\rm eff}$ and $\mu$ is presented for $m_X=[10,1000]$~MeV and $\tau_X=[10^3,10^{10}]$~s. Then we show  two slices of the $\Delta N_{\rm eff}$ and $\mu$ predictions by fixing $m_X=100$~MeV (left panel) and $\tau_X=10^7$~s (right panel), respectively. The three curves in both panels correspond to different initial $Y_X$ abundances, following the simple scaling $\Delta N_{\rm eff}\propto Y_X, \mu\propto Y_X^2$. In the left panel, the plateau of the curves corresponds to the $\mu$ distortion formation era, with the arrow indicating the increasing of lifetimes from $10^3$~s to $10^{10}$~s. In the right panel, both $\Delta N_{\rm eff}$ and $\mu$ increase as the  decaying particle mass becomes larger. From both panels, we can draw a general conclusion that for low-energy neutrino injection with $1~\text{MeV}<E_\nu<1$~GeV, the $\Delta N_{\rm eff}$  prediction will exceed the current observational bounds well before the $\mu$ distortion can reach the forecast detection limit $|\mu|=10^{-8}$, even if the pair annihilation channel  is more important than coannihilation in the low-energy regime. Therefore,  joint observations of an $N_{\rm eff}$ excess and a primordial CMB $\mu$ distortion are not attainable in this energy injection regime. Note that, however, this conclusion may be changed if there is a larger pair annihilation rate mediated by light hidden particles, such as $\nu+\bar\nu\to Z'\to e^++e^-$. This interesting possibility deserves consideration elsewhere, which may target the correlation of $\Delta N_{\rm eff}$ and CMB spectral distortions at $1~\text{MeV}<E_\nu<1$~GeV.

\begin{figure*}[t]
	\centering
	\includegraphics[scale=0.43]{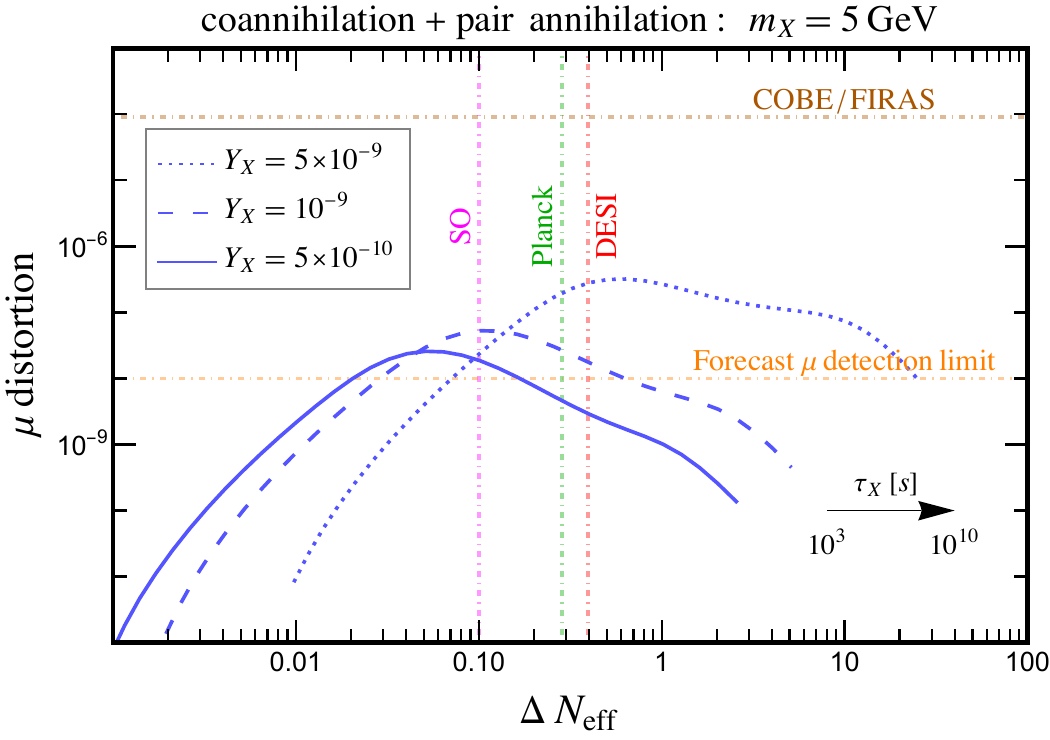} \quad
	\includegraphics[scale=0.432]{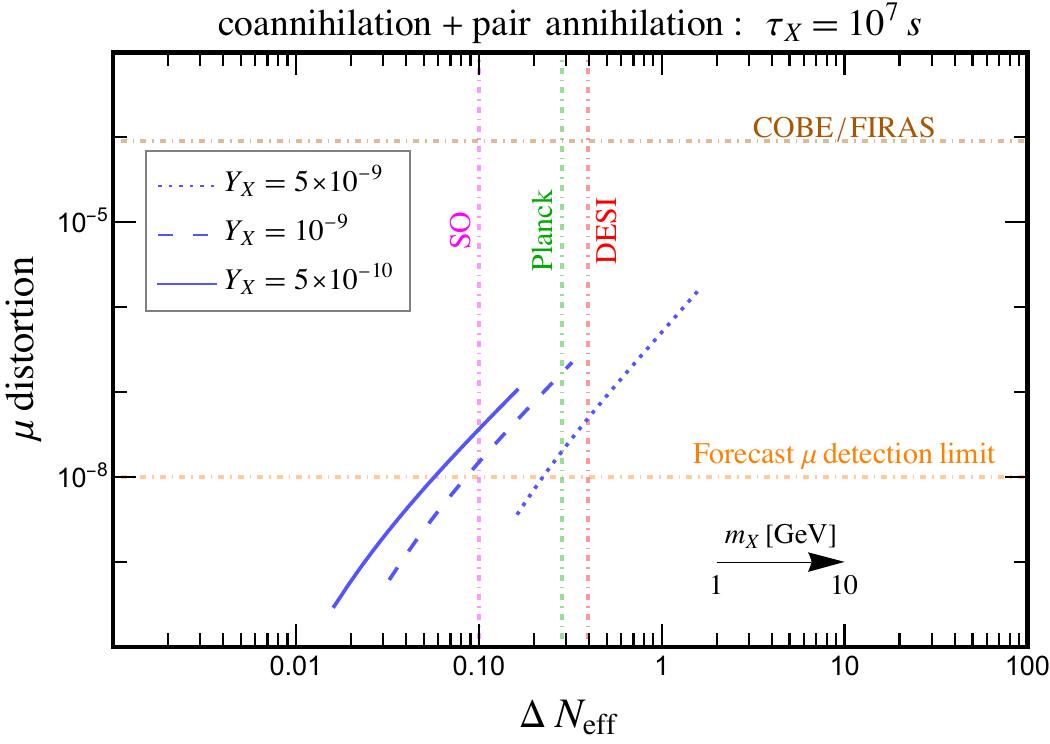} 
	\caption{\label{fig:Neffmu-tot}The correlated predictions of $\Delta N_{\rm eff}$ and the CMB $\mu$ distortion from pair annihilation and coannihilation  of injected neutrinos $\nu_{\rm inj}+\bar\nu_{\rm inj}\to e^++e^-, \nu_{\rm inj}+\bar\nu_{\rm bg} (\bar\nu_{\rm inj}+\nu_{\rm bg}) \to e^++e^-$. See the caption of Fig.~\ref{fig:Neffmu-pair} for more details. 
	}
\end{figure*}

We show in Fig.~\ref{fig:Neffmu-tot} the correlation between $\Delta N_{\rm eff}$ and the $\mu$ distortion  which is induced from both pair annihilation and coannihilation. Taking the slice of $m_X=5$~GeV as an example, we see from the left panel that joint observations of  $\Delta N_{\rm eff}$ and   $\mu$ are possible. This can be realized with an initial abundance $Y_X<10^{-8}$. Note that for $\tau_X>10^3$~s, $Y_X=10^{-8}$ corresponds to the upper bound derived from BBN observations~\cite{Kanzaki:2007pd,Chang:2024mvg,Bianco:2025boy}.  This indicates that joint observations of $\Delta N_{\rm eff}$ and $\mu$ can probe the parameter space that has weak impacts on BBN. In particular, it can probe a long-lived particle with a much smaller initial abundance at decay.  From the left panel, we can also infer that the maximal $\mu$ distortion can reach $\mu=\mathcal{O}(10^{-7})$, predicting $\Delta N_{\rm eff}$ within the current detection region at the same time. This is qualitatively consistent with the expectation given by Eq.~\eqref{eq:Deltarho_gamma-rho_gamma-1}.  In the right panel, we show the slice of $\tau_X=10^7$~s, where the bottom (top) edge of the curves corresponds to $m_X=1 (10)$~GeV. From these panels, we numerically find that $m_X>1$~GeV and $\tau_X>10^4$~s are generally  required to create the joint observational windows for $\Delta N_{\rm eff}$ and $\mu$.

In general, we confirm the picture that there is a large overlap of parameter space in which an excess of $N_{\rm eff}$ at current and future detection sensitivities, $0.01-0.1$,  will  be accompanied with $\mu\gtrsim 10^{-8}$ that is observable in forecast detection limits.  This picture holds even with an initial abundance of the long-lived particles much smaller than the upper bounds derived from BBN. Note that a lifetime $\tau_X>10^4$~s indicates that neutrino injection occurs at $T\lesssim 0.01$~MeV,  which is  later than  neutron-proton freeze-out~\cite{Pitrou:2018cgg}. On the other hand, neutrino injection may also lead to depletion of light nuclei formed during BBN via photo-disintegration or hadro-disintegration. Nevertheless, these disintegration effects in general require ultrahigh-energy neutrino injection to produce secondary photons or hadrons. In addition, disintegration effects also depend on the injected neutrino number density. As found in Refs.~\cite{Chang:2024mvg,Bianco:2025boy}, for $\tau_X>10^4$~s and 1~GeV~$<E_\nu<100$~GeV, the upper bounds on $Y_X$ from BBN disintegration effects are typically at $\mathcal{O}(10^{-8}-10^{-7})$. For smaller $Y_X$, dominant constraints will be derived by $N_{\rm eff}$. This strengthens the perspective that the synergy of $N_{\rm eff}$ and CMB spectral distortions can play an important role in uncovering the underlying source for the non-electromagnetic or non-hadronic energy injection without creating significant impacts on BBN. 

Finally, let us comment on the differences between   lower and higher neutrino energy injection.  For higher energy injection  with  $E_\nu>100$~GeV, $\Delta N_{\rm eff}$ could be either positive or negative, since a large photon energy will  be generated from electroweak cascade and contribute to a negative $\Delta N_{\rm eff}$ well before $z\simeq 2\times 10^6$ via photon temperature shift~\citep{Chluba:2020oip}. Similarly,  it was noticed in Ref.~\cite{Acharya:2020gfh} that constraints from CMB spectral distortions  can supersede those from $N_{\rm eff}$  if the injected neutrino energy is much higher $E_\nu\gtrsim 100$~GeV. In particular, the bounds from current CMB spectral distortions can forbid a large   $\Delta N_{\rm eff}$ generation in the regime $0.01-0.1$. This implies that there would be less feasibility to search for the underlying injection source via the synergy of $N_{\rm eff}$ and CMB spectral distortions from ultrahigh-energy neutrino injection. For lower-energy neutrino injection, as we have shown here, the synergy is readily visible without significant, complicated electromagnetic or electroweak cascades.

\section{Conclusion}
Several non-electromagnetic energy injection after neutrino decoupling can readily yield the same prediction of $\Delta N_{\rm eff}$, implying a large degeneracy of this observable in probing the underlying physics origin. In this work, we proposed that observations of CMB spectral distortions can be combined with  that from $\Delta N_{\rm eff}$, even if the energy injection is in the form of non-electromagnetic species such as neutrinos and neutral dark radiation. We demonstrated this property by considering a kind of neutrinogenic CMB spectral distortions, where extra neutrino energy injection occurs in the early Universe and subsequently gives rise to a small electromagnetic energy leakage via neutrino annihilation. We found that a positive shift of $\Delta N_{\rm eff}\simeq 0.01-0.1$ that reaches the current and future detection sensitivities can be accompanied by a CMB $\mu$ distortion reaching the forecast detection regime $\mu\gtrsim 10^{-8}$. In terms of long-lived particle decay, the joint observations of $\Delta N_{\rm eff}$ and the CMB $\mu$ distortion can be realized by a particle mass above 1~GeV with a lifetime longer than $10^4$~s, where the initial particle abundance can be much smaller than from BBN constraints. This work thus highlights a new synergy between CMB anisotropy experiments and absolute CMB spectroscopy, as may become available in the future.

\section*{Acknowledgements}
We would like to thank John Beacom for helpful discussions on neutrino injection from dark matter annihilation, which may also lead to CMB spectral distortions. S.-P.~Li is supported by JSPS Grant-in-Aid for JSPS Research Fellows No. 24KF0060.

\appendix

\section{Analytic estimate of neutrino annihilation}
\label{app:rate}
The total electromagnetic energy transfer from neutrino coannihilation reads
\begin{align}
    \frac{\Delta \rho_{\gamma, \rm co}}{\rho_\gamma}&\equiv \int_{0}^{\infty} \mathcal{J}_{\rm dis}(T)\frac{{\rm d} \rho_{\gamma,\rm co}/{\rm d}t}{\rho_\gamma HT}{\rm d}T\label{eq:Drho-co-approx}
    \\[0.2cm]
    &\simeq \frac{G_{\rm F}^2 g_W}{3\pi}\int_{0}^{\infty}\mathcal{J}_{\rm dis}(T)\mathcal{J}_{\rm inj} (T)\frac{{\rm d} T}{\rho_\gamma H T}
      \nonumber\\[0.2cm]
      &\times \int\frac{{\rm d}^3 {\bf p}_\nu}{(2\pi)^3 2E_\nu}\frac{{\rm d}^3 {\bf p}_{\bar\nu}}{(2\pi)^3 2E_{\bar \nu}} s^2 (E_\nu+E_{\bar\nu})f_\nu f_{\bar \nu}
         \nonumber\\[0.2cm]
    &\simeq\frac{G_{\rm F}^2 g_W}{3\pi}\int_{T_{f}}^{T_i} \frac{E_\nu \Delta \rho_\nu \rho_{\bar \nu}}{\rho_\gamma H T}{\rm d} T
           \nonumber\\[0.2cm]
    &\simeq\frac{7G_{\rm F}^2 g_W}{24\pi}\left(\frac{4}{11}\right)^{4/3}\Delta N_{\rm eff} E_\nu \int_{T_{f}}^{T_i} \frac{\rho_{\bar \nu}}{H T}{\rm d} T\,.     \nonumber
\end{align}
In  the second line, we  use the Heaviside step-function for the visibility function of distortions, such that $\mathcal{J}_{\rm dis}(T)=\mathcal{J}_\mu(T)=\theta(T_\mu-T)\theta(T-T_{\mu y})$ for the $\mu$ distortion and $\mathcal{J}_y(T)=\theta(T_{\mu y}-T)$ for the $y$ distortion~\cite{Chluba:2016bvg},  and focus on the energy transfer at  $T_{\mu y}\lesssim T_f<
T_i\lesssim T_\mu$ for the $\mu$ distortion formation and $T_f<
T_i\lesssim T_{\mu y}$ for the $y$ distortion formation, where $\mathcal{J}_{\rm inj}(T)=\theta(T_i-T)\theta(T-T_{f})$ is introduced, with $T_i, T_f$ denoting the initial and final moments of the transfer.  In addition, we approximate $\mathcal{G}\approx g_W$ in Eq.~\eqref{eq:G}, and neglect the energy threshold in Eq.~\eqref{eq:ee-int}.  We should then keep in mind that this approximation is not valid for $E_\nu$ below the $e^\pm$ production threshold. Besides, we neglect the angular term such that $s\approx 2p_\nu\cdot p_{\bar \nu}\simeq 2E_\nu E_{\bar \nu}$. In the third line, we take the limit $E_\nu+E_{\bar\nu}\approx E_\nu$ and use
\begin{align}
\int\frac{{\rm d}^3 {\bf p}_\nu}{(2\pi)^3} E_\nu^2 f_\nu&\sim E_\nu \Delta \rho_\nu\,,
\\[0.1cm]
\int\frac{{\rm d}^3 {\bf p}_{\bar\nu}}{(2\pi)^3} E_{\bar\nu} f_{\bar \nu}&=\rho_{\bar \nu}=\frac{7}{8}\frac{\pi^2}{30}T_\nu^4\,.
\end{align}
The first equation assumes a monochromatic energy spectrum and neglects the redshift effect of the injected neutrino energy, while the second equation takes the thermal background neutrino distribution with $T_\nu\sim T$ being the neutrino temperature.  To derive  the last line of Eq.~\eqref{eq:Drho-co-approx}, we use Eq.~\eqref{eq:DeltaNeff-nu} to express $\Delta \rho_\nu$ in terms of a constant $\Delta N_{\rm eff}$. Finally, the temperature integration yields
\begin{align}
\int_{T_{f}}^{T_i} \frac{\rho_{\bar \nu}}{H T}{\rm d} T\propto T_i^2-T_f^2\,,
\end{align}
giving rise to Eq.~\eqref{eq:Deltarho_gamma-rho_gamma-1}.

The total electromagnetic energy transfer from neutrino  pair annihilation reads
\begin{align}
    \frac{\Delta \rho_{\gamma, \rm  pair}}{\rho_\gamma}&\equiv \int_{0}^{\infty} \mathcal{J}_{\rm dis}(T)\mathcal{J}_{\rm inj} (T)\frac{{\rm d} \rho_{\gamma,\rm pair}/{\rm d}t}{\rho_\gamma HT}{\rm d}T \label{eq:Drho-pair-approx}
\\[0.1cm]
    &\simeq\frac{G_{\rm F}^2 g_W}{3\pi}\int_{T_{f}}^{T_i} \frac{E_\nu \Delta \rho_\nu^2 }{\rho_\gamma H T}{\rm d} T
           \nonumber\\[0.1cm]
    &\simeq\frac{49 G_{\rm F}^2 g_W}{768\pi}\left(\frac{4}{11}\right)^{8/3}\Delta N_{\rm eff}^2 E_\nu \int_{T_{f}}^{T_i} \frac{\rho_{\gamma}}{H T}{\rm d} T \,,\nonumber
\end{align}
where similar approximations to derive Eq.~\eqref{eq:Drho-co-approx} have also been applied here, but with $E_\nu+E_{\bar\nu}\approx 2 E_\nu$ for pair annihilation. Again, we take the monochromatic energy spectrum without the redshift effect
\begin{align}
\int\frac{{\rm d}^3 {\bf p}_{\nu(\bar\nu)}}{(2\pi)^3} E_{\nu(\bar\nu)} f_{\nu(\bar\nu)}&\sim  \Delta \rho_{\nu}=\Delta \rho_{\bar\nu}\,.
\end{align}
Then we use Eq.~\eqref{eq:DeltaNeff-nu} to express $\Delta \rho_\nu$ in terms of a constant $\Delta N_{\rm eff}$, noting that $\Delta \rho_\nu$ accounts for half of $\Delta N_{\rm eff}$ from pair annihilation. Finally, integrating over the temperature of Eq.~\eqref{eq:Drho-pair-approx} will yield Eq.~\eqref{eq:Deltarho_gamma-rho_gamma-2}.

\bibliographystyle{utphys}
\bibliography{Refs}

\end{document}